\documentclass{aa}  

\usepackage{graphicx}
\usepackage{txfonts}
\usepackage[pdfpagelabels=false]{hyperref}	% Hyperlinks
\hypersetup{colorlinks=true,linkcolor=blue,citecolor=blue,filecolor=blue,urlcolor=blue,}
\newcommand{\pcm}{~cm$^{-2}$}	% per cm-squared
\newcommand{\pcmm}{~cm$^{-3}$}	% per cm-cubed
\newcommand{\kms}{~km\,s$^{-1}$}
\newcommand{\Hii}{H~{\sc ii}}

\usepackage{xcolor}
\newcommand{\iz}[1]{{\color{black} #1}}

\begin{document}

   \title{Fine structure and kinematics of the ionized and molecular gas in the jet and disk around S255IR NIRS3 from high resolution ALMA observations}
   \titlerunning{Ionized and molecular gas in the jet and disk around S255IR NIRS3}

%   \subtitle{I. Overviewing the $\kappa$-mechanism}

   \author{I. I. Zinchenko\inst{1}
          \and
         S.-Y. Liu\inst{2}
          \and
         Y.-N. Su\inst{2}
          }

   \institute{Federal Research Center A.V. Gaponov-Grekhov Institute of Applied Physics of the Russian Academy of Sciences, \\ 46 Ul’yanov str., Nizhny Novgorod 603950, Russia\\
              \email{zin@iapras.ru}
         \and
             Institute of Astronomy and Astrophysics, Academia Sinica, 11F of ASMAB, AS/NTU No.1, Sec. 4, Roosevelt Rd, Taipei 10617, Taiwan\\
             \email{syliu@asiaa.sinica.edu.tw, ynsu@asiaa.sinica.edu.tw}
%             \thanks{The university of heaven temporarily does not                     accept e-mails}
             }

   \date{Received XXX; accepted YYY}

% \abstract{}{}{}{}{} 
% 5 {} token are mandatory
 
  \abstract
   {}
   {We present observations of the high-mass star-forming region S255IR, which harbors the $\sim$20~M\sun\ protostar NIRS3, where a disk-mediated accretion burst was recorded several years ago, with the angular resolution of $\sim$15~mas, which corresponds to $\sim$25~au and is almost an order of magnitude better than in the previous studies of this object.}
   {The observations were performed with ALMA \iz{at a wavelength} of 0.9~mm in continuum and in several molecular lines.}
   {In \iz{the} continuum we detected the central bright source (brightness temperature $\sim$850~K) elongated along the jet direction and two pairs of bright knots in the jet lobes. These pairs of knots imply a double ejection from NIRS3 with the time interval of $\sim$1.5~years. The orientation of the jet differs by $\sim$20$^\circ$ from that on larger scales, as mentioned also in some other recent works. The 0.9~mm continuum emission of the central source represents a mixture of the dust thermal emission and free-free emission of the ionized gas. Properties of the free-free emission are typical for hypercompact \Hii\ regions. In the continuum emission of the knots in the jet the free-free component apparently dominates. In the molecular lines a sub-Keplerian disk around NIRS3 about 400~au in diameter is observed. The absorption features in the molecular lines towards the central bright source may indicate an infall. The molecular line emission \iz{appears} very inhomogeneous at small scales, which \iz{may} indicate a small-scale clumpiness in the disk.}
   {}

   \keywords{Stars: formation  -- Stars: massive -- ISM: clouds -- ISM: molecules -- ISM: individual objects (S255IR) -- Submillimeter: ISM
               }

   \maketitle
%
%-------------------------------------------------------------------

\section{Introduction}

There are still several competing scenarios of high-mass star formation \citep[e.g.,][]{Tan14, Motte18, Rosen2020, Padoan2020}. In recent years, a large attention has been paid to the luminosity outbursts in massive protostars, which are believed to be caused by episodic disk-mediated accretion events. There are theoretical models which predict such a behavior \citep[e.g.,][]{Meyer17, Meyer19}. To date, several such bursts have been recorded \citep{Caratti16, Brogan2019, Proven-Adzri2019, Hunter17, Tapia2013, Chen2021, Wolf2024}. One of the first of them was the burst in S255IR NIRS3, which was observed at IR \citep{Caratti16} and submillimeter \citep{Liu18} wavelengths, and was accompanied by the methanol maser flare \citep{Moscadelli17, Szymczak2018}.

S255IR at the distance of $1.78_{-0.11}^{+0.12}$~kpc \citep{Burns16} is a well-known site of high-mass star formation \citep[e.g.,][]{Zinchenko2024vak}, a part of the large star-forming complex sandwiched between the evolved \Hii\ regions S255 and S257 \citep{Ojha2011}. It contains three major cores SMA1, SMA2 and SMA3 \citep{Wang11} and several smaller condensations \citep{Zin2020}. The SMA1 core harbors a $\sim$20~M\sun\ protostar NIRS3 \citep{Zin15}. The mass is estimated from the bolometric luminosity of $\sim 3\times 10^4$~L\sun\ at the adopted distance.

A disk-outflow system is associated with this protostar \citep{Zin15}. Assuming Keplerian rotation, from the SMA \iz{(SubMillimeter Array)} data \citet{Zin15} derived the inclination angle of the disk of $\sim$25$^\circ$. However, the IR image implies the disk seen almost edge-on \citep{Boley13}. An analysis of the ALMA data with a much higher angular resolution of $\sim$0{\farcs}14 results in the conclusion that the observed rotating structure represents an infalling envelope or pseudo-disk in sub-Keplerian rotation \citep{Liu2020}. \citet{Liu2020} assumed an upper limit for the radius of the probable centrifugal barrier at $\sim$125~AU.

The IR jet in this area was detected many years ago \citep{Howard97}. Recently multi-frequency (up to $\sim$92~GHz) detailed studies of the radio jets were performed with the highest angular resolution of $\sim$0{\farcs}1 \citep{Obonyo2021, Cesaroni2023, Cesaroni2024}. 

Here we report and discuss new ALMA observations of this object with the angular resolution almost an order of magnitude higher than in the previous studies.

\section{Observations} \label{sec:obs}

We carried out our observations with the ALMA toward S255IR SMA1 under the project \#2019.1.00315.S to image the continuum emission at the highest angular resolution of $\sim$0{\farcs}015 ($\sim$ 27 au). One execution was carried out on 2021 September 3rd at a hybrid C43-9/10 configuration with 45 antennas in the 12-m array. The on-source integration time of the C43-9/10 execution is about 18.5 mins, and the baseline lengths range from 121 m to 16.2 km. The phase center in the ICRS reference frame was R.A. = 06$^\mathrm{h}$12$^\mathrm{h}$54{\fs}013 (J2000) and decl. = $+$17\degr59\arcmin23{\farcs}050. To \iz{maximize} the aggregated bandwidth for the continuum maps, the ALMA correlator was configured to provide four 1.875 GHz spectral windows centered at 334.593, 336.583, 346.593, and 348.593 GHz with a spectral resolution of 1.13 MHz ($\sim$1.0 km s$^{-1}$) after the online Hanning-smooth. Molecular line transitions such as C$^{34}$S, SiO (8$-$7), CO (3$-$2) and CH$_3$CN (19$-$18) are simultaneously captured by the designed spectral windows. The half-power width of the ALMA 12-m primary beam was $\sim$17$.04\arcsec$ at 341.5 GHz, more than sufficient to cover the whole extents of the S255IR SMA1. The quasar J0510+1800 was observed to calibrate passband and absolute flux scale. The quasar J0613+1708 was observed as the gain calibrator.  The acquired visibility data were reduced using the observatory pipeline in CASA (Common Astronomy Software Application) version 6.2.1.7. The synthesized beam of the continuum map was $\sim 19\times 13$~mas, which corresponds to $\sim 34\times 23$~AU at the S255IR distance. The synthesized beams for the spectral line maps are practically the same.

Under the project \#2019.1.00315.S, we also carried out band 7 observations at $\sim$0{\farcs}1 in order to monitor the temporal variation of S255IR SMA1 in dust continuum and molecular line
emission. Two executions were carried out on 2021 June 13th and July 6th with 41 and 40, respectively, antennas in the 12-m array. The total on-source integration time is about 90 mins. The synthesized beam was $\sim 0.109\times 0.975$~arcsec.

\section{Results} \label{sec:results}

\subsection{Continuum emission} \label{sec:cont}

\begin{figure}
    \centering
    \includegraphics[width=\columnwidth]{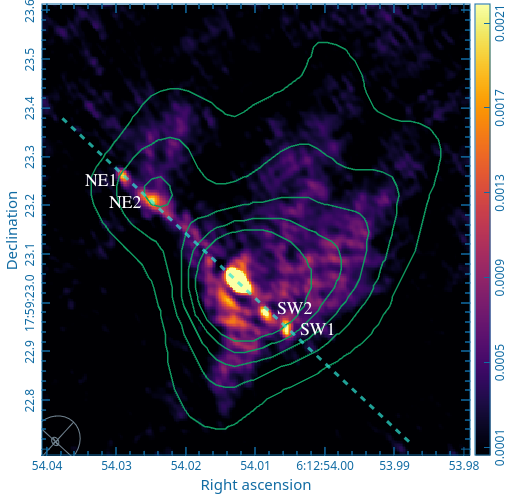}
    \caption{The continuum image of the S255IR region. The intensity scale is in Jy\,beam$^{-1}$. The contours show the low resolution continuum emission. The contour levels are from 4 to 20 in step of 4~mJy\,beam$^{-1}$. The dashed line has the position angle of 47$^\circ$. The four bright knots in the jet lobes are marked.}
    \label{fig:cont}
\end{figure}

In Fig.~\ref{fig:cont} we present the continuum image of the S255IR region. It shows several compact bright features and a fainter extended structure elongated in the same way as the disk (or pseudo-disk) observed here earlier at lower resolutions \citep{Zin15, Liu2020}.

The brightest feature in the center practically coincides with the nominal NIRS3 position. The peak brightness of this feature is approximately 20~mJy\,beam$^{-1}$ (Table~\ref{tab:cont}), which corresponds to about 850~K. The total flux density of this feature is approximately 60~mJy integrated in the circle of 45~mas in radius or about 50~mJy from the 2D Gaussian fit (Table~\ref{tab:cont}). Its size from the 2D Gaussian fitting is 0{\farcs}033$\times$0{\farcs}019, which corresponds to $\sim$60$\times$34~AU$^2$. The position angle is PA $\approx 41^\circ$.

There are 4 other bright knots in the image (NE1, NE2, SW1 and SW2), located almost on a strait line along with the central source, two on each side from this source. Properties of these knots are summarized in Table~\ref{tab:cont}. Their peak brightness is from 2.0 to 2.7~mJy\,beam$^{-1}$ ($\sim$80--110~K) and fluxes $\sim$4--8~mJy. The position angle of the strait line which connects all these knots is approximately 47$^\circ$. Apparently, these knots belong to the jet emanating from the central source. It is worth noting that the central source is elongated approximately in the direction of the jet.

\begin{table*}
 \caption{Properties of the continuum clumps obtained from the 2D Gaussian fitting.}
 \label{tab:cont}
 \small
 \begin{tabular}{lccccccc}
  \hline\hline
  Object & R.A.(J2000) & Dec.(J2000) &B &$\theta_\mathrm{max}$ &$\theta_\mathrm{min}$ &P.A. &Flux \\
   & $^\mathrm{h\;m\;s}$  & $^{\circ\; \prime\; \prime\prime}$ &mJy\,beam$^{-1}$ &mas &mas &deg &mJy\\
  \hline
  Central & 6:12:54.01263 ± 0.00001 & 17:59:23.0500 ± 0.0002 &19.7 ± 0.3 &33.2 ± 0.6 &19.0 ± 0.2 &41.1 ± 0.9 &50.5 ± 1.1\\
  NE1 & 6:12:54.02893 ± 0.00007 & 17:59:23.2618 ± 0.0016 &2.3 ± 0.3 &24.6 ± 4.2 &14.6 ± 1.6 &--151 ± 8 &3.3 ± 0.7\\
  NE2 & 6:12:54.02472 ± 0.00018 & 17:59:23.2125 ± 0.0021 &2.0 ± 0.3 &49.5 ± 7.4 &28.1 ± 3.5 & 57 ± 9 &11.5 ± 1.8\\
  SW1 & 6:12:54.00545 ± 0.00009 & 17:59:22.9527 ± 0.0017 &2.4 ± 0.3 &29.2 ± 4.1 &23.7 ± 3.0 &--20 ± 24 &6.9 ± 1.2\\
  SW2 & 6:12:54.00849 ± 0.00010 & 17:59:22.9815 ± 0.0020 &2.5 ± 0.3 &37.6 ± 5.4 &20.7 ± 2.2 & 33 ± 9 &7.8 ± 1.3\\
  \hline
 \end{tabular}
 \normalsize
\end{table*}

%The fainter extended emission looks very clumpy but regular patterns can be seen in this clumpiness. Therefore, the inhomogeneity can be, at least partly, an artifact related to the interferometer filtering and data processing. The peak brightness of this emission is $\sim$1~mJy\,beam$^{-1}$.

\subsection{Line emission} \label{sec:lines}
The observed bands cover many molecular lines. \iz{Here,} we present observational results for several strongest and the most informative lines, which help \iz{in} understanding the structure, kinematics and physical properties of this region.

\subsubsection{C$^{34}$S(7--6)}
\begin{figure*}
    \centering
    \includegraphics[width=0.95\textwidth]{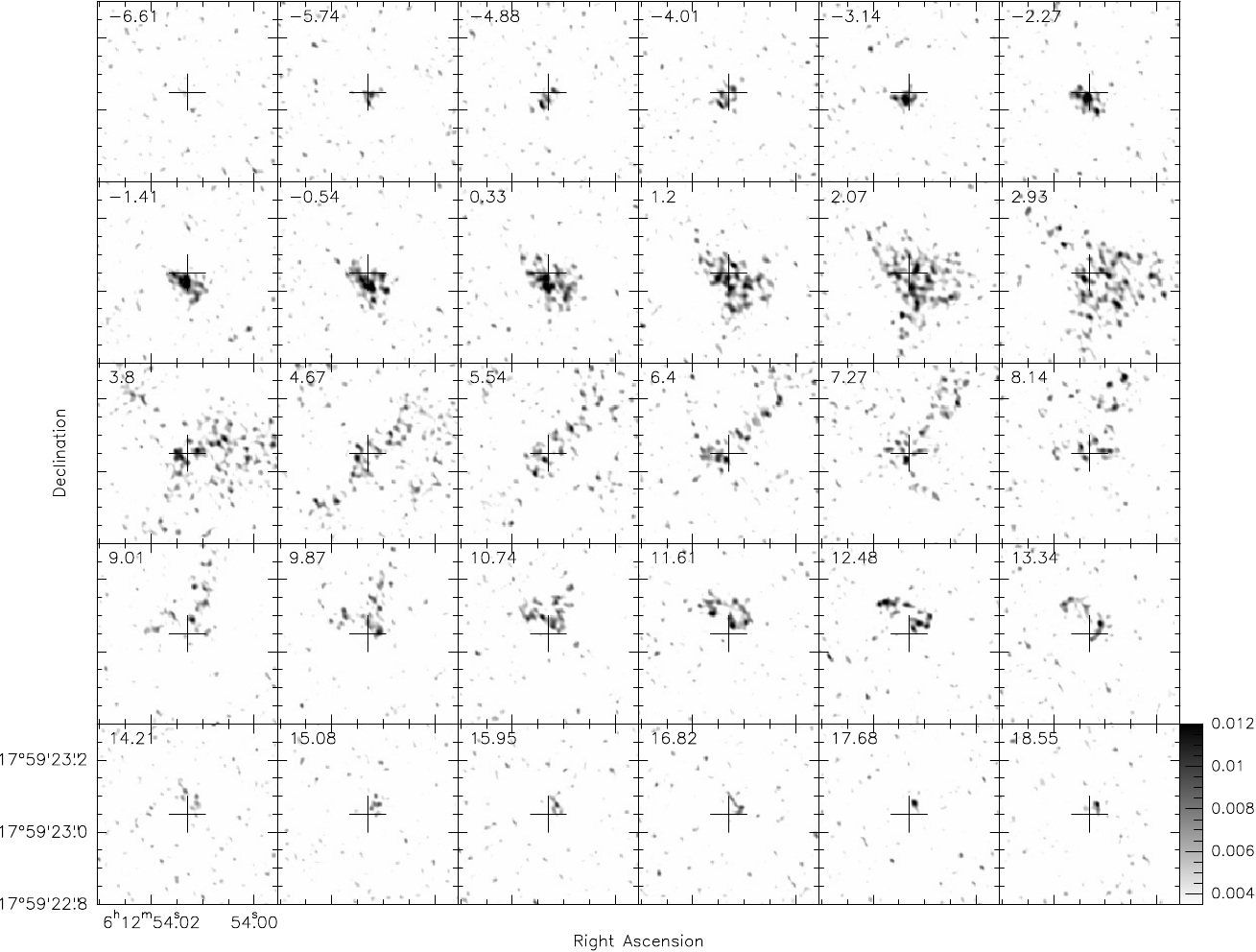}
    \caption{The channel maps of the C$^{34}$S(7--6) emission in the the S255IR region. The intensity scale is in Jy\,beam$^{-1}$. The cross marks the phase center of the observations.}
    \label{fig:c34s-chmap}
\end{figure*}

%\subsubsection{Emission maps and PV diagrams}
In Fig.~\ref{fig:c34s-chmap} we present the channel maps of the C$^{34}$S(7--6) emission.

\begin{figure*}
    \centering
    \includegraphics[width=0.33\textwidth]{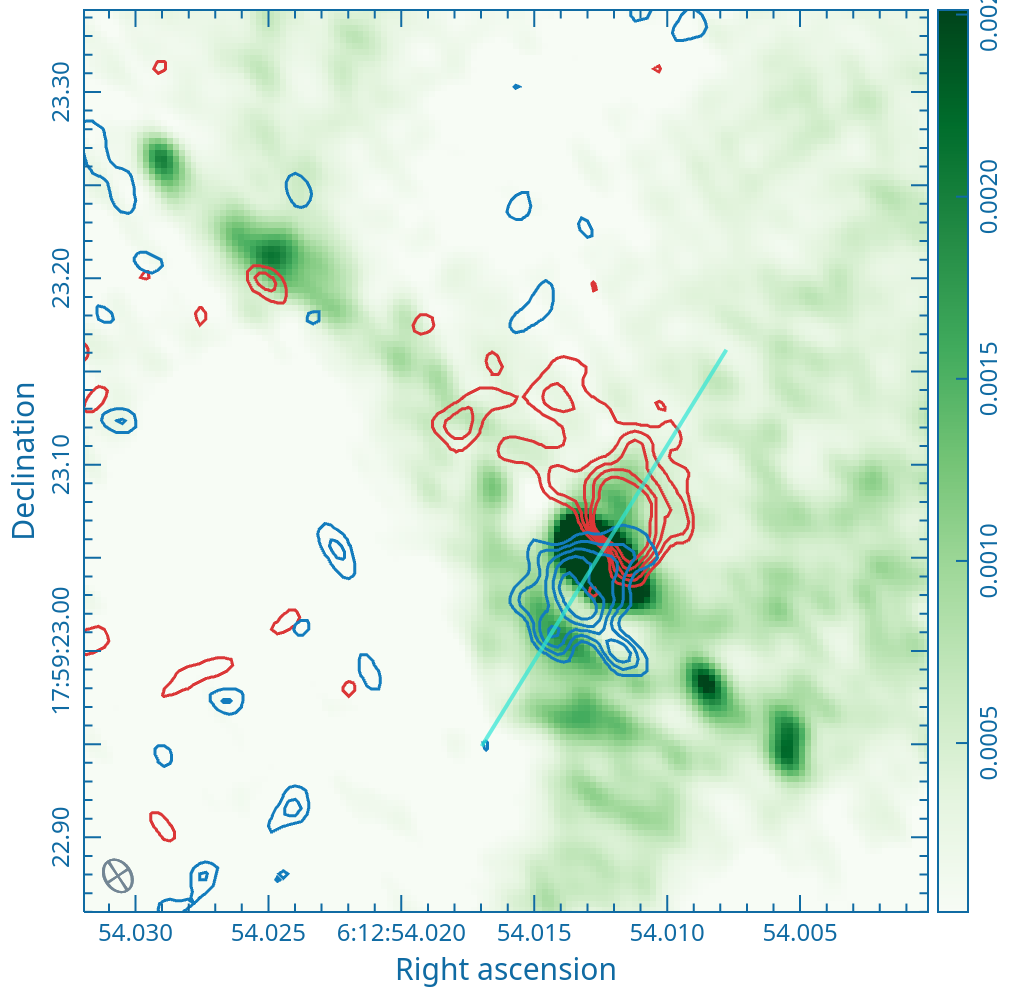}
    \hfill
    \includegraphics[width=0.33\textwidth]{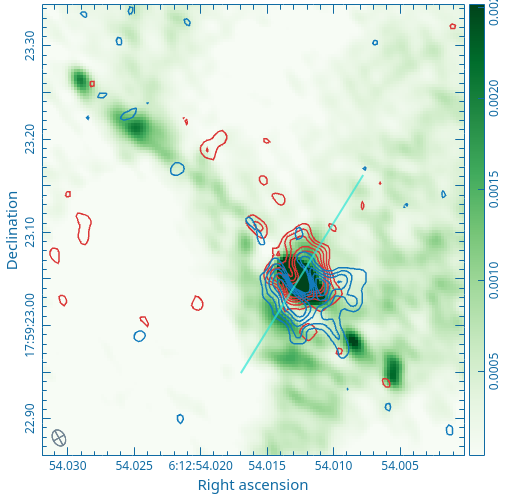}
    \hfill
    \includegraphics[width=0.33\textwidth]{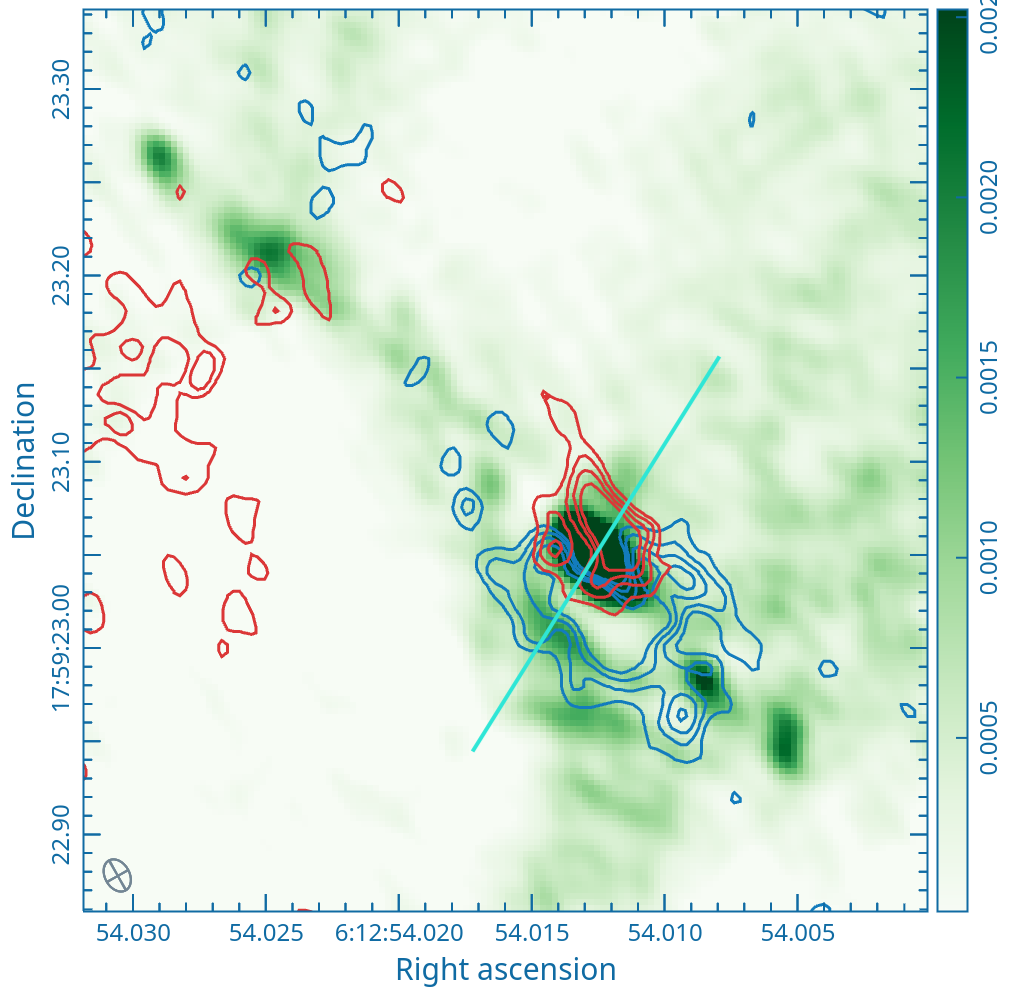}
    \caption{Contours of the blue-shifted and red-shifted emission in the C$^{34}$S(7--6) (left panel), SiO(8--7) (central panel) and CO(3--2) (right panel) line emission overlaid on the continuum image. The intensity scale is in Jy\,beam$^{-1}$. The C$^{34}$S(7--6) blue emission is integrated from --15 to --2\kms, the C$^{34}$S(7--6) red emission is integrated from 10 to 25\kms. The contour levels are 12, 19, 26, 33 and 40~mJy\,beam$^{-1}$ for the blue-shifted emission and 15.0; 23.8; 32.5; 41.3 and 50.0~mJy\,beam$^{-1}$ for the red-shifted emission. The SiO(8--7) blue emission is integrated from --20 to 0\kms, the SiO(8--7) red emission is integrated from 10 to 30\kms. The contour levels are 24, 38, 52, 66 and 80~mJy\,beam$^{-1}$ for both blue-shifted and red-shifted emission. The CO(3--2) blue emission is integrated from --30 to --10\kms, the CO(3--2) red emission is integrated from 15 to 40\kms. The contour levels are 28.5, 39, 49.5 and 60~mJy\,beam$^{-1}$ for the blue-shifted emission and 38, 52, 66 and 80~mJy\,beam$^{-1}$ for the red-shifted emission. 
    The cyan line shows the position of the PV cut (PA = 328$^\circ$).}
    \label{fig:wings}
\end{figure*}

The emission \iz{appears} very clumpy. %This clumpiness does not look \iz{artificial} and is most probably real. 
The velocity gradient is clearly seen in these maps. In Fig.~\ref{fig:wings} we present maps of the blue-shifted and red-shifted C$^{34}$S(7--6) emission overlaid on the continuum image.

\begin{figure*}
    \centering
    \includegraphics[width=0.33\textwidth]{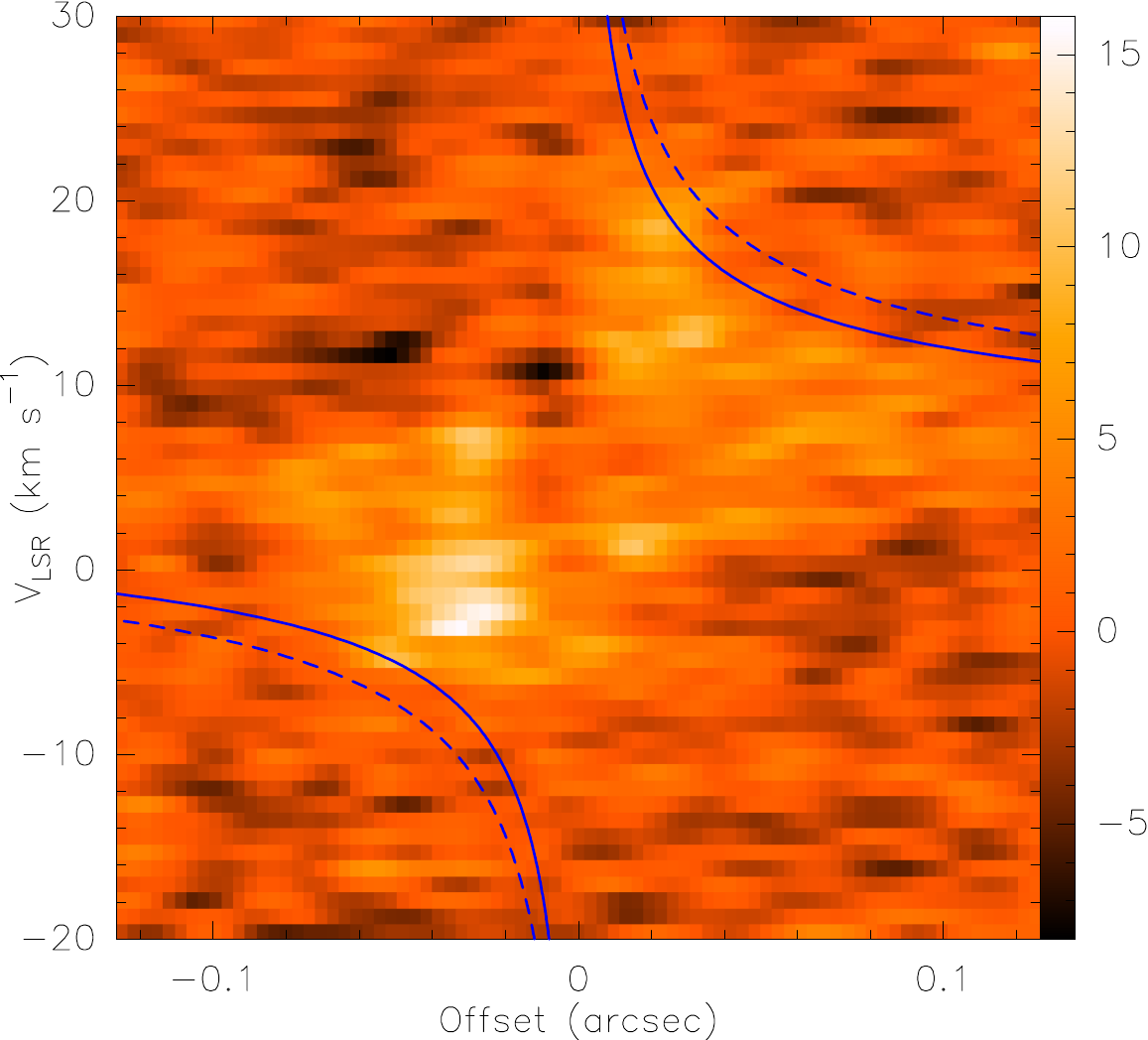}
    \hfill
    \includegraphics[width=0.33\textwidth]{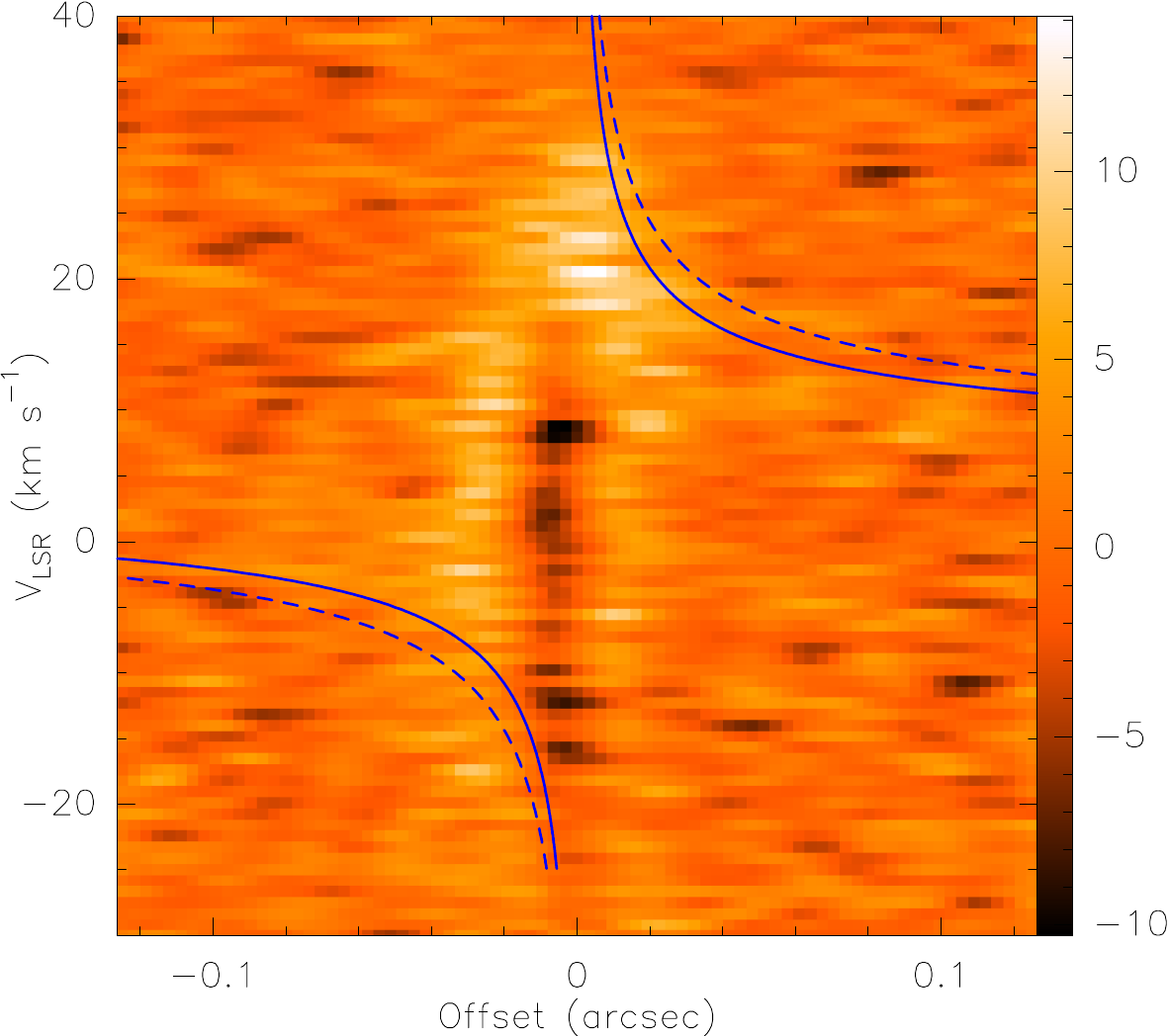}
    \hfill
    \includegraphics[width=0.33\textwidth]{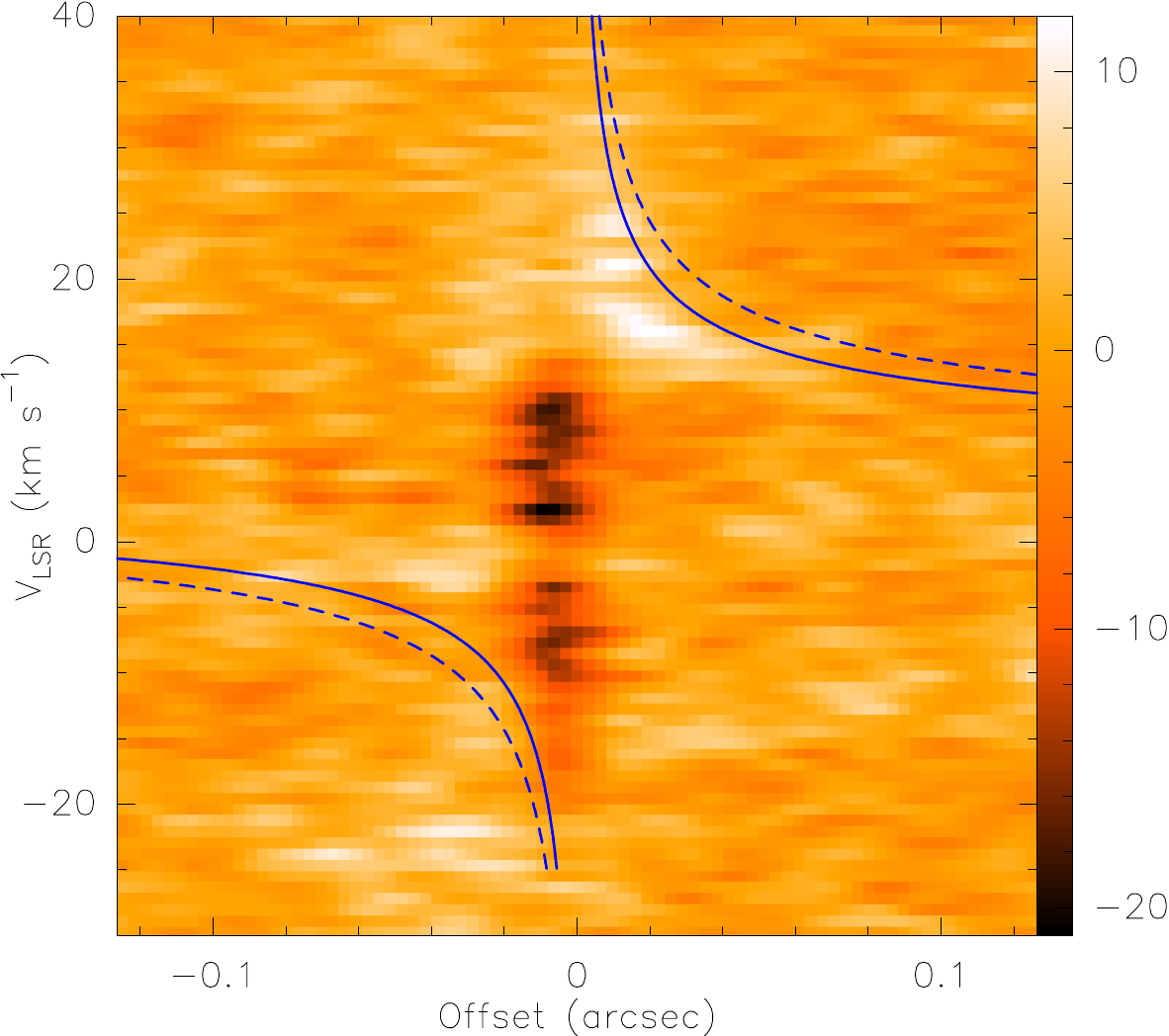}
    \caption{The position-velocity diagrams for the C$^{34}$S(7--6) (left panel),  SiO(8--7) (central panel) and CO(3--2) (right panel) emission generated along the path shown in Fig.~\ref{fig:wings}. The intensity scale is in mJy\,beam$^{-1}$. The curves correspond to Keplerian rotation around the central mass of $M \sin^2i = 10$~M\sun\ (solid) and $M \sin^2i = \iz{15}$~M\sun\ (dashed), where $i$ is the inclination angle.}
    \label{fig:pv}
\end{figure*}

The C$^{34}$S(7--6) emission in Fig.~\ref{fig:wings} clearly shows an elongated (almost perpendicular to the jet) and apparently rotating structure. Most probably it represents a disk-like object. Its extension along the major axis is about 400~AU. We generated a position-velocity (PV) diagram along this structure (the PV path is shown in Fig.~\ref{fig:wings}). It is presented in Fig.~\ref{fig:pv} (left panel). The curves correspond to Keplerian rotation around the central mass of $M \sin^2i = 10$~M\sun\ (solid) and $M \sin^2i = \iz{15}$~M\sun\ (dashed), where $i$ is the inclination angle. It is worth noting that the C$^{34}$S(7--6) emission extends also along the jet and might indicate a velocity gradient across the jet (Fig.~\ref{fig:wings}).

\subsubsection{SiO(8--7)}
The morphology of the SiO(8--7) emission is similar to that of the C$^{34}$S(7--6) emission (Fig.~\ref{fig:wings}). It only seems to be somewhat more compact along the disk and more extended along the blue-shifted outflow lobe. The PV diagram in the SiO(8--7) line is presented in Fig.~\ref{fig:pv} (central panel). In this diagram there is a prominent, broad in velocity negative feature toward the central position, discussed \iz{in Subsection~\ref{sec:abs}}.

\subsubsection{CO(3--2)}
The CO(3--2) emission in the vicinity of the central source (Fig.~\ref{fig:wings}, right panel) shows about the same morphology as C$^{34}$S(7--6) and SiO(8--7). However it is much more extended along the outflow lobes. An especially extended and bright emission is observed toward the red-shifted lobe. The CO(3--2) emission along the continuum jet in vicinity of the central source may indicate a rotation. The PV diagram for the CO(3--2) emission along the same path as for C$^{34}$S(7--6) and SiO(8--7) is shown in Fig.~\ref{fig:pv} (right panel). It also shows a strong negative feature toward the central position.

\subsubsection{CH$_{3}$CN(19--18)}
The CH$_{3}$CN(19--18) emission in this area is relatively weak at this high angular resolution and is confined to several compacts regions. The strongest emission is observed in the disk and near the NE2 and SW2 knots in the jet. %In Fig.~\ref{fig:ch3cn_spec} we present the spectrum of the CH$_{3}$CN(19--18) emission averaged in the circle of 0{\farcs}08 in radius around the central source. The first 10 components of the $K$-ladder are clearly seen.

%\begin{figure}
%    \centering
%    \includegraphics[width=\columnwidth]{figs/ch3cn_k3_mom0}
%    \caption{The map of the integrated intensity in the CH$_{3}$CN(19$_3$--18$_3$) line overlaid with contours of the 0.9~mm continuum emission. The intensity scale is in Jy\,beam$^{-1}$\,\kms. The contour levels are 0.3, 0.6, 0.9, 1.2 and 1.5~mJy\,beam$^{-1}$.}
%    \label{fig:ch3cn_map}
%\end{figure}

%\begin{figure}
%    \centering
%    \includegraphics[width=\columnwidth]{figs/ch3cn_spec}
%    \caption{The CH$_{3}$CN(19--18) spectrum averaged in the circle of 0{\farcs}08 in radius around the central source. The $K$ values of the components are indicated. The strong line at 349.1~GHz is the CH$_3$OH $14_1 - 14_0$~A$^{-+}$ maser line \citep{Zin17}.}
%    \label{fig:ch3cn_spec}
%\end{figure}

\subsubsection{The methanol maser lines}
Earlier we detected a new methanol maser line ($14_1 - 14_0$~A$^{-+}$ at 349.1~GHz) in this area \citep{Zin17}. It is seen in the new data, too. Now, there was one more line of this series in the observed bands, $12_1 - 12_0$~A$^{-+}$. It also shows the maser effect. We will present and discuss these results elsewhere. The maser effect in the lines of this series can be a tracer of luminosity flares during high-mass star formation \citep{Salii2022}.

\subsubsection{Absorption features} \label{sec:abs}
\begin{figure}
    \centering
    \includegraphics[width=\columnwidth]{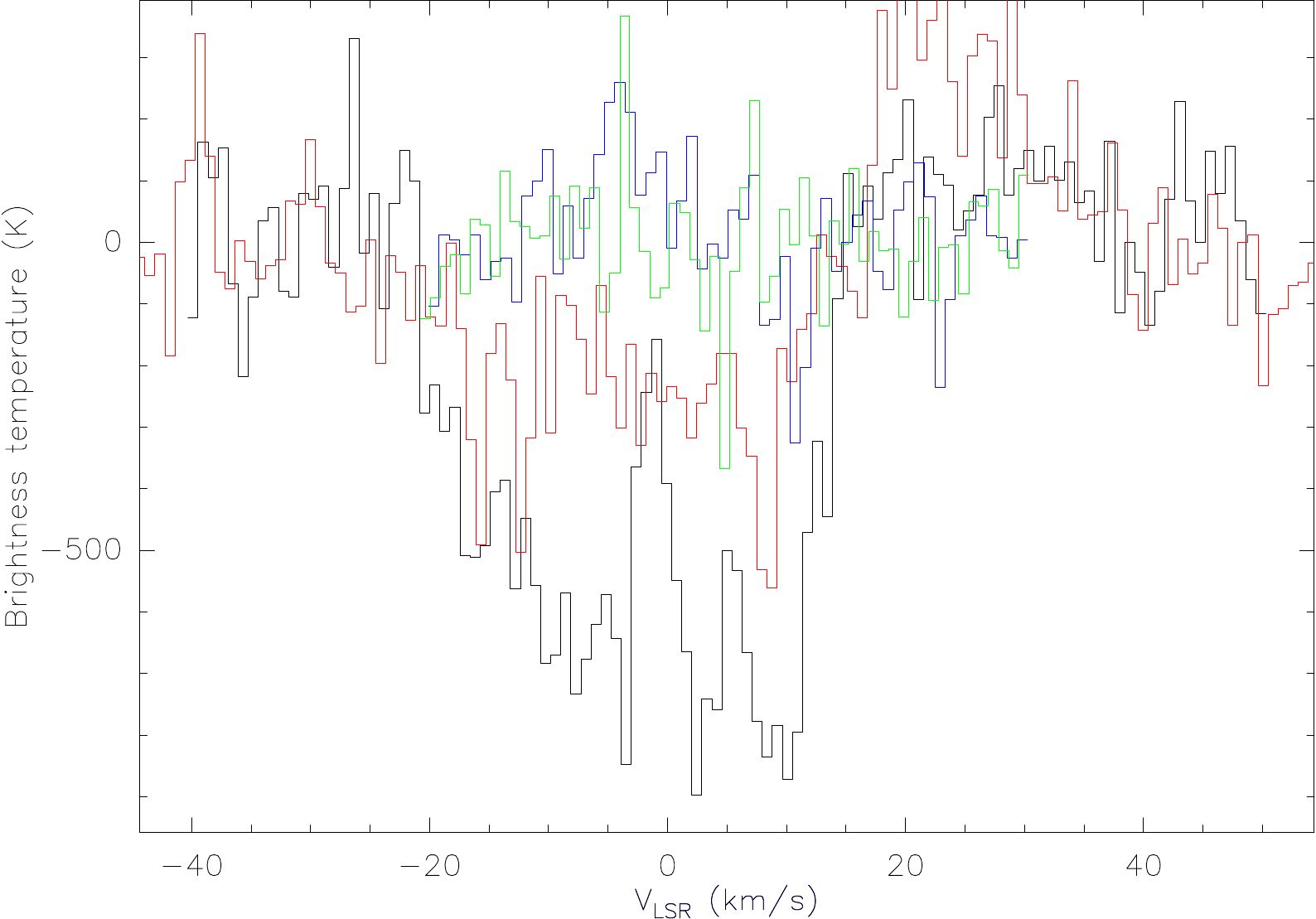}
    \caption{The spectra of CO(3--2) (black), SiO(8--7) (red), C$^{34}$S(7--6) (blue) and CH$_{3}$CN(19$_3$--18$_3$) (green) toward the central source (R.A. = 6$^\mathrm{h}$12$^\mathrm{m}$54{\fs}0125, Dec. = 17$^\circ$59$^\prime$23{\farcs}053).}
    \label{fig:absorption}
\end{figure}

As noted above, strong and wide absorption features are observed in the SiO(8--7) and CO(3--2) lines toward the central source. \iz{In Fig.~\ref{fig:absorption}, we present the spectra of all the lines discussed earlier at this central position}. An absorption, although much more narrow, is observed also in C$^{34}$S(7--6) and \iz{perhaps in} CH$_{3}$CN(19--18).

The deepest absorption in the C$^{34}$S(7--6), SiO(8--7) and CO(3--2) lines is seen at $V_\mathrm{LSR}\sim 10$\kms. In the CO(3--2) line it is about --800~K. %In the CH$_{3}$CN(19$_3$--18$_3$) line the absorption feature, if real, is observed at $\sim$5\kms. 
At the same time there are emission features at about --5\kms\ in C$^{34}$S(7--6) and CH$_{3}$CN(19$_3$--18$_3$), and at about 20\kms\ in CO(3--2) and SiO(8--7).

\section{Discussion} \label{sec:discussion}

\subsection{General morphology}
The results presented above show very clearly a rotating disk-like structure around the NIRS3 protostar and the jet originating in this structure. It is necessary to note that the orientation of this jet is significantly different from that observed at larger scales. According to the rather old IR data the position angle of the jet axis is $\approx 67^\circ$ \citep{Howard97}. A similar orientation has been observed in several other works \citep{Zin15, Cesaroni18, Obonyo2021}. It differs by $\approx 20^\circ$ from that displayed in Fig.~\ref{fig:cont}. However, recent observations at small scales show the jet orientation similar to that reported above. \citet{Cesaroni2023} found the position angle of the jet of 48$^\circ$, which practically coincides with our result. A very similar orientation was found by \citet{Hirota2021} from observations of water masers. These results indicate the disk precession as suggested in some previous works \citep{Obonyo2021, Cesaroni2023}. It is worth noting that this precession can results not only in the change of the position angle of the outflow but also in the change of the inclination angle. This inclination angle is believed to be about 80$^\circ$, i.e. the disk is seen practically edge-on \citep{Boley13}. However, it can change by about the same amount as the position angle. It is not excluded that the large opening angle of the molecular outflow from the SMA1 clump \citep{Zin15, Zin2020} can be explained by these variations of the jet orientation. The 
\iz{orientation of the} extended red-shifted CO(3--2) emission (Fig.~\ref{fig:wings}) %is consistent with wide opening angle outflow observed earlier and its orientation 
is closer to the position angle \iz{of the outflow} observed at large scales.

Our data show two pairs of bright knots in the jet, one in the each lobe. These pairs of knots were observed in the same areas as the NE and SW knots in \citet{Cesaroni2023} in 2021, i.e. in about the same time as our observations. In those observations these pairs of knots were seen as single entities, apparently due to a much lower angular resolution. According to our data the projected distance between the NE knots is 138~AU and between the SW knots it is 93~AU. The projected expansion speed is estimated as $\sim$450\kms\ for the NE lobe \citep{Fedriani2023, Cesaroni2023}  and $\sim$285\kms\ for the SW lobe \citep{Cesaroni2024}. Therefore, the observed distances between the knots correspond to about 530 days for the NE pair of knots and to about 570 days for the SW pair. Most probably, the presence of these two pairs of knots implies two events of ejection from the central source with the time interval of $\sim$550 days or 1.5 years. It may imply two events of accretion, respectively. An inspection of the 6.7~GHz methanol maser \citep{Szymczak2018} and NIR \citep{Uchiyama2020} light curves show their complex behavior during the two years after the original burst. The light curve of the 6.7~GHz methanol maser at 2.87\kms\ has two peaks with the same time interval between them \citep{Szymczak2018}, \iz{which is consistent with the assumption of two ejection events}. %An additional burst is not excluded.

\subsection{Nature of the continuum emission}

\subsubsection{The central source}
The high brightness of the central source as well as its morphology (an elongation along the jet axis) hint at the free-free emission of the ionized gas. Of course, the dust contribution is not excluded, too. \citet{Liu2020} found the dust emission to be optically thin with the brightness temperature of $\sim$120~K at the 0{\farcs}14 scale. In principle the dust column density and brightness can be higher at higher resolution. On the other hand the contribution of the free-free emission to the measurements at the lower resolution should be non-negligible. The deep absorption in the molecular lines (Fig.~\ref{fig:absorption}) favors a bright compact central source.

%Assuming the dominant contribution of the free-free emission, we can estimate first its optical depth. Under the assumption of a typical electron temperature in \Hii\ region of 7000--10000~K, the optical depth is $\sim$0.1. Then, using the well-known formulae for free-free emission \citep[e.g.,][]{Wilson2013} we obtain the emission measure of $EM\sim 5\times 10^{10}$~pc\,cm$^{-6}$. For the source size of $\sim$40~AU it implies the electron density of $n_\mathrm{e}\sim 10^7$\pcmm. The turnover frequency for these parameters is $\sim$100~GHz. Such properties are typical for hypercompact \Hii\ regions \citep{Kurtz2005}.

\begin{figure}
    \centering
    \includegraphics[width=\columnwidth]{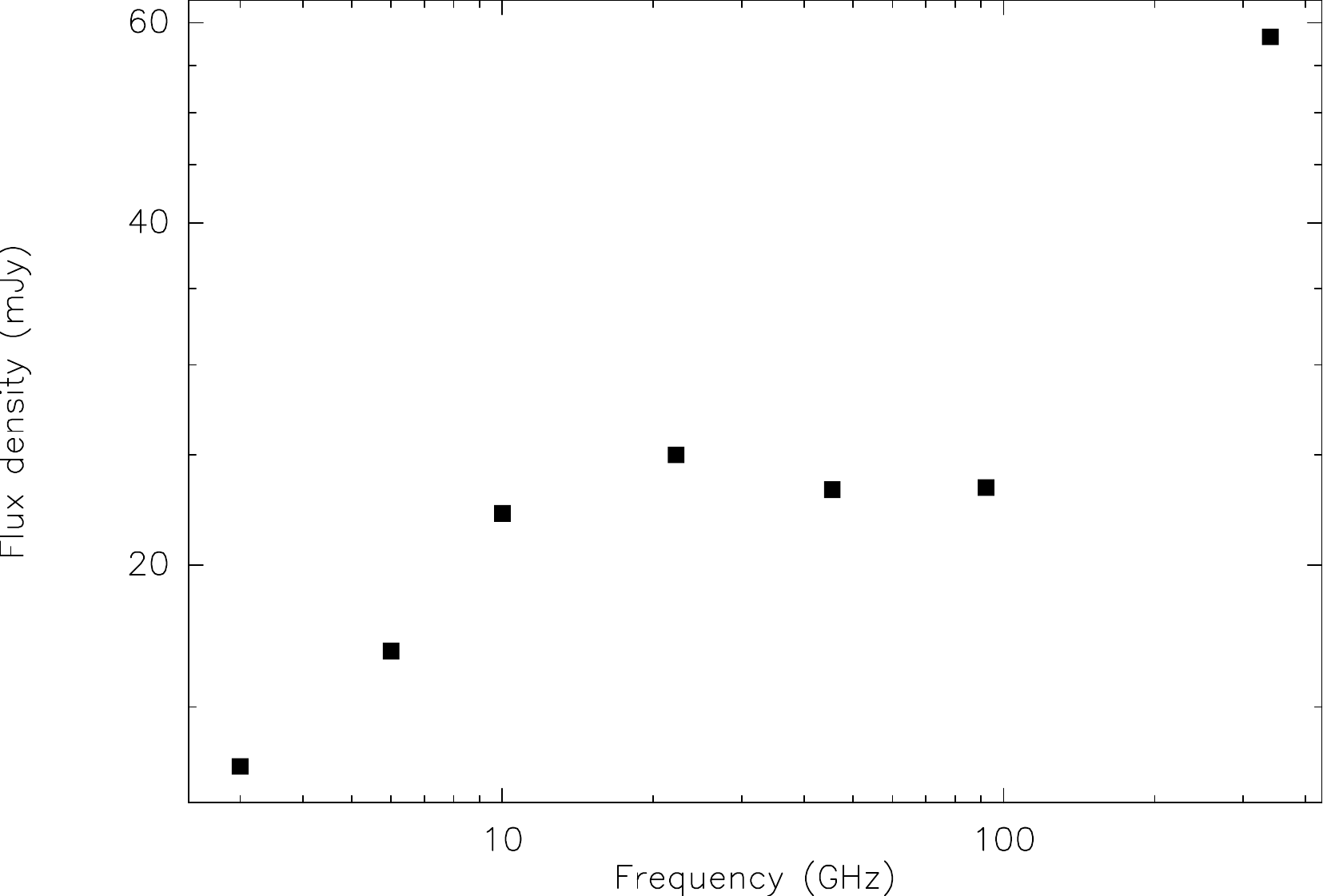}
    \caption{The measured fluxes of S255 NIRS3 from \citet{Cesaroni2023} (from 3 to 92.2~GHz) and this work (340~GHz). At 340~GHz a summary flux of the central clump and the SW2 knot is plotted.}
    \label{fig:nirs-fluxes}
\end{figure}

\citet{Cesaroni2023} reported the flux densities for S255IR NIRS3 at several frequencies measured at different epochs. The latest measurements at most frequencies were performed in 2018. The data at 92.2~GHz were obtained in September 2021, at about the same time as our observations. The beam size in 2018 measurements was from 2{\farcs}3 at 3~GHz to 0{\farcs}2 at 45.5~GHz. In the measurements at 92.2~GHz the synthesized beam was 0{\farcs}087. Such a beam covers also the SW2 knot. In our data, the flux integrated in the circle of 0{\farcs}09 in diameter is $\sim$60~mJy. However, it can include a significant contribution from the surrounding dust emission. \citet{Obonyo2021} reported fluxes for this object in 2018 of 24.72 ± 0.80~mJy at 22~GHz and 15.11 ± 0.09~mJy at 6~GHz, very similar to those measured by \citet{Cesaroni2023}. The beam sizes were about 0{\farcs}1 and 0{\farcs}28 at 22~GHz and 6~GHz, respectively. In Fig.~\ref{fig:nirs-fluxes} we plot the fluxes reported by \citet{Cesaroni2023} and our summary flux of the central clump and the SW2 knot. It is necessary to bear in mind that the measurements have been made in different times and with different beams. 
%Nevertheless, this spectrum is consistent with the emission of an ionized stellar envelope. The slope of the spectrum at frequencies \la 100~GHz is $\sim$0.4, which does not contradict such an interpretation.

The spectrum in Fig.~\ref{fig:nirs-fluxes} looks very typical for a mixture of the free-free and dust emission. 
%\iz{There are several methods, which are used for a disentanglement of their contributions \citep[e.g.,][]{Purser2021}. However, they are hardly directly applicable here for the following reasons. First, the measurements discussed have been done at different times and with different beams, which insert significant uncertainties. Then, these methods assume a simple summation of the dust and free-free contributions, which is valid only in the case of low optical depth for both components. For the dust emission this is probably not true, as shown below.} 
\iz{Assuming an optically thin free-free emission with the spectral index of --0.1 at frequencies {\ga}22~GHz, we obtain from the measurements at 22.2 and 45.5~GHz \citep{Cesaroni2023} the free-free flux at 340~GHz of about 19~mJy. This value can be considered as an upper limit for the optically thin free-free emission of the ionized gas in the central source because the measurements in \citet{Cesaroni2023} refer to area including this source and the SW lobe. %Moreover, the emission maps at different frequencies in \citet{Cesaroni2024} show that this emission is mostly associated with the jet lobe. 
Our summary flux for the knots in the SW lobe is $14.7\pm 1.8$~mJy (Table~\ref{tab:cont}). If it is fully attributed to the free-free emission, the share of the central source will be only $\sim$4~mJy. The map of the free-free emission at 92.2~GHz in \citet{Cesaroni2024} (their Figure 3b) shows the brightness at the central source position of $\sim$8~mJy\,beam$^{-1}$. Therefore, for further estimates we can consider the range of the free-free flux from the central source, assuming an optically thin emission, as $\sim$4--19~mJy from 50.5~mJy (Table~\ref{tab:cont}). The preferale value can be $\sim$7~mJy, as follows from the measurements at 92.2~GHz at the same epoch as our observations.}

\iz{It should be noted that \citet{Cesaroni2024} obtained that only 14.6~mJy from 23.5~mJy at 92.2~GHz in this area refer to the free-free emission, while the remaining 8.9~mJy they attribute to the dust emission. In this case the spectral index of this emission (attributed to dust) between 92 and 340~GHz will be $\sim$1.5, which is not consistent with the optically thin dust emission and even with the black body emission. We consider possible reasons for this discrepancy below.} 

%In this interpretation about 21~mJy at 340~GHz could be attributed to the free-free emission and about 37~mJy to dust. 
\iz{In the framework of the model of the optically thin free-free emission of the ionized gas and dust emission,}
in terms of brightness temperature, \iz{assuming the same spatial distribution of both components, the obtained estimates of the fluxes} imply the brightness of the free-free emission of $\sim$\iz{70--320}~K and brightness of the dust emission of $\sim$\iz{530--780}~K (in our beam). However, such a simple summation, would be valid only in the case of low optical depth for both components. The optical depth for the free-free emission in this picture is small, but for dust it is not so. The value of \iz{even $\sim$530}~K is much higher than the brightness of the dust emission in the 0{\farcs}14 beam \citep{Liu2020} and implies a very high dust temperature. The upper limit for the dust temperature is the sublimation temperature, which is $\sim$1500--2000~K for graphite and $\sim$1000--1500~K for silicates \citep{Churchwell1990}. Such a temperature can be reached at the inner boundary of the dust cocoon around a massive star. This implies a substantial dust optical depth. At the same time the optical depth of the central core cannot be too high because we see emission features in the spectra in Fig.~\ref{fig:absorption}. These emission features should arise in the gas located behind the central core. For the dust optical depth of $\sim$1, the dust temperature should be $\ga$\iz{850}~K. This is a very high value but probably not excluded in vicinity of the massive YSO with the luminosity of a few $10^4$~L\sun. Under the same assumptions about the dust opacity as in \citet{Liu2020} the dust optical depth of $\tau\sim 1$ implies the hydrogen column density of $N\sim \iz{2}\times 10^{25}$\pcm\ and the volume density of $n\sim 3\times 10^{10}$\pcmm. It is worth noting that \citet{Liu2020} derived the gas density of $\sim 6.6 \times 10^9$\pcmm\ in the 0{\farcs}14 beam. However,  if we consider the value of dust absorption coefficient for ``naked'' dust at a density of $10^8$\pcmm\ from \citet{Ossenkopf94} the estimates of the hydrogen column and volume densities will decrease by an order of magnitude.

The brightness of the free-free emission of $\sim$\iz{70--320}~K (unaffected by the dust absorption) implies the optical depth of $\sim$\iz{0.007--}0.03 for the typical gas temperature of $\sim$10000~K. Then, using the well-known formulae for free-free emission \citep[e.g.,][]{Wilson2013} we obtain the emission measure of $EM\sim \iz{(0.5-2)}\times 10^{10}$~pc\,cm$^{-6}$. For the source size of $\sim$40~AU it implies the electron density of $n_\mathrm{e}\sim\iz{(0.5-1)}\times 10^7$\pcmm. Such properties are typical for hypercompact \Hii\ regions \citep{Kurtz2005}. For a uniform mixture of gas and dust these estimates imply a very low ionization fraction. It is not excluded that dust, at least partly, is expelled from the ionized region and forms a dusty cocoon around it. \iz{When the size of the ionized gas emission region is smaller than the size of the dust emission region, the relative brightness of the free-free emission in the center will be higher.}

\iz{An alternative model for this source can be based on a hypercompact \Hii\ region with the turnover frequency \ga200~GHz, which can provide the spectral index of $\sim$1.5, derived above for the excess emission in the frequency range 92--340~GHz, due to decrease of the electron density with radius \citep{Kurtz2005}. It implies the emission measure of $EM\ga 2\times 10^{11}$~pc\,cm$^{-6}$. Such a model resolves the problem with the spectral index in the range 92--340~GHz but it predicts a too high brightness temperature, taking into account the observed source size (since the optical depth at 340~GHz will be rather high). The solution can include a dust absorption. 
Observations at several frequencies with a sufficiently high resolution are required to select the most appropriate model.}

\subsubsection{Knots in the jet}
\begin{figure*}
    \centering
    \includegraphics[width=0.4\textwidth]{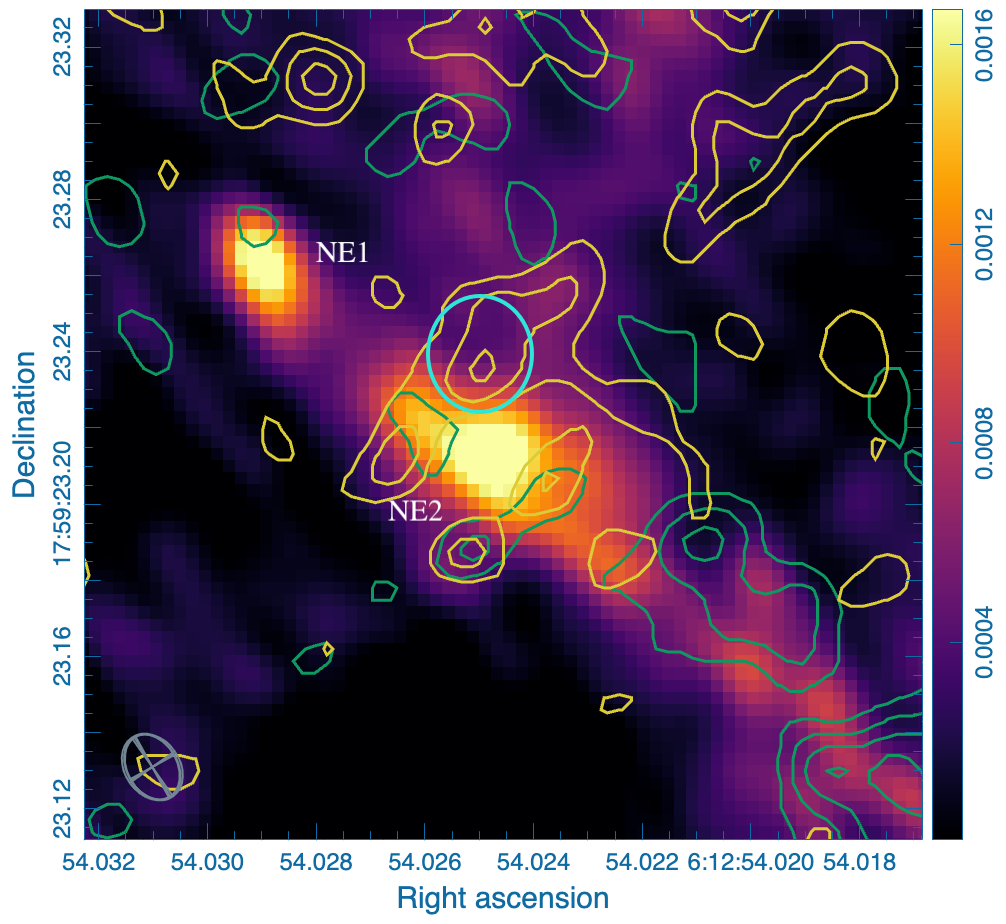}
    \hspace{5mm}
    \includegraphics[width=0.4\textwidth]{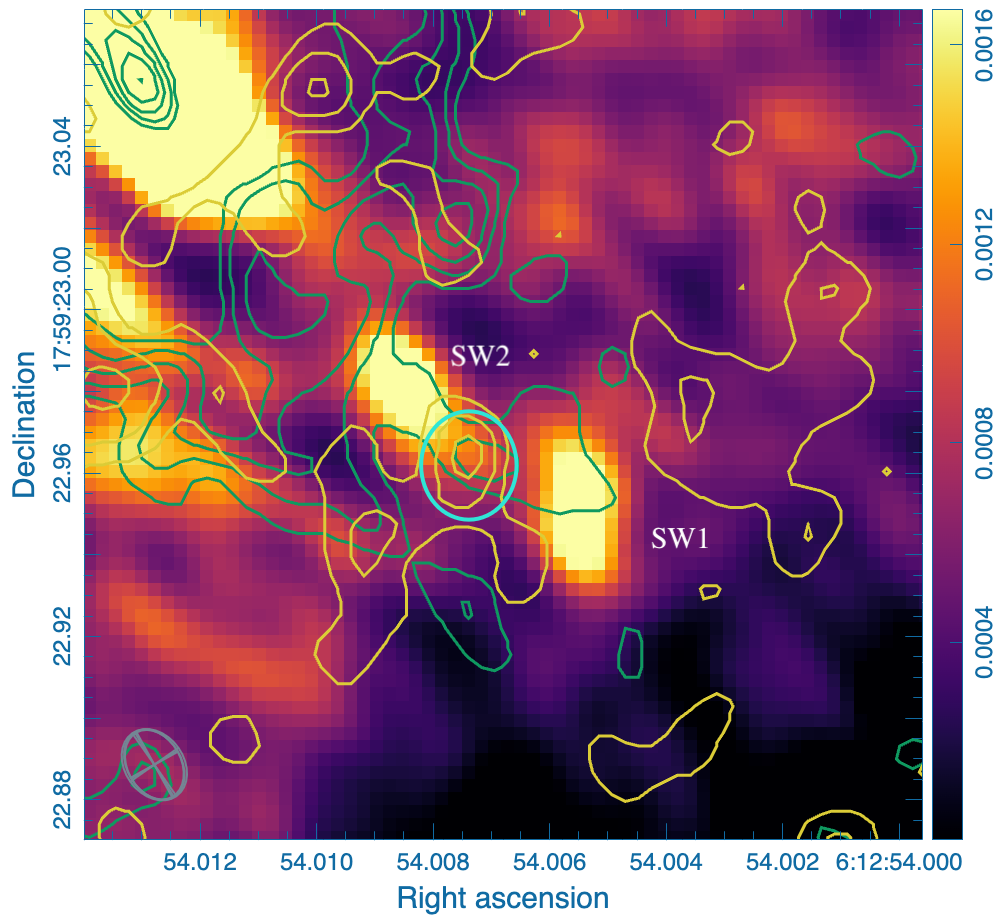}
    \caption{Zoomed-in images of the NE (left panel) and SW (right panel) jet lobes in continuum at 0.9~mm overlaid with contours of the C$^{34}$S(7--6) (green) and CH$_{3}$CN(19$_3$--18$_3$) (yellow) integrated line intensity. The contour levels are 12.5, 21.9, 31.3, 40.6 and 50~mJy\,beam$^{-1}$\,\kms\ for both C$^{34}$S(7--6) and CH$_{3}$CN(19$_3$--18$_3$). The cyan ovals show the regions where the C$^{34}$S(7--6) and CH$_{3}$CN(19$_3$--18$_3$) spectra presented in Figure~\ref{fig:knots_spectra} were extracted.}
    \label{fig:knots_maps}
\end{figure*}

Now, let us consider the emission of the knots in the jet. In Figure~\ref{fig:knots_maps} we present zoomed-in views of the NE and SW knots in continuum as well as in the C$^{34}$S(7--6) and CH$_{3}$CN(19$_3$--18$_3$) lines. There is no molecular emission associated with the NE1 and SW1 knots, while emission peaks in these lines are observed in vicinity of the NE2 and SW2 knots. 

\begin{figure*}
    \centering
    \includegraphics[width=0.4\textwidth]{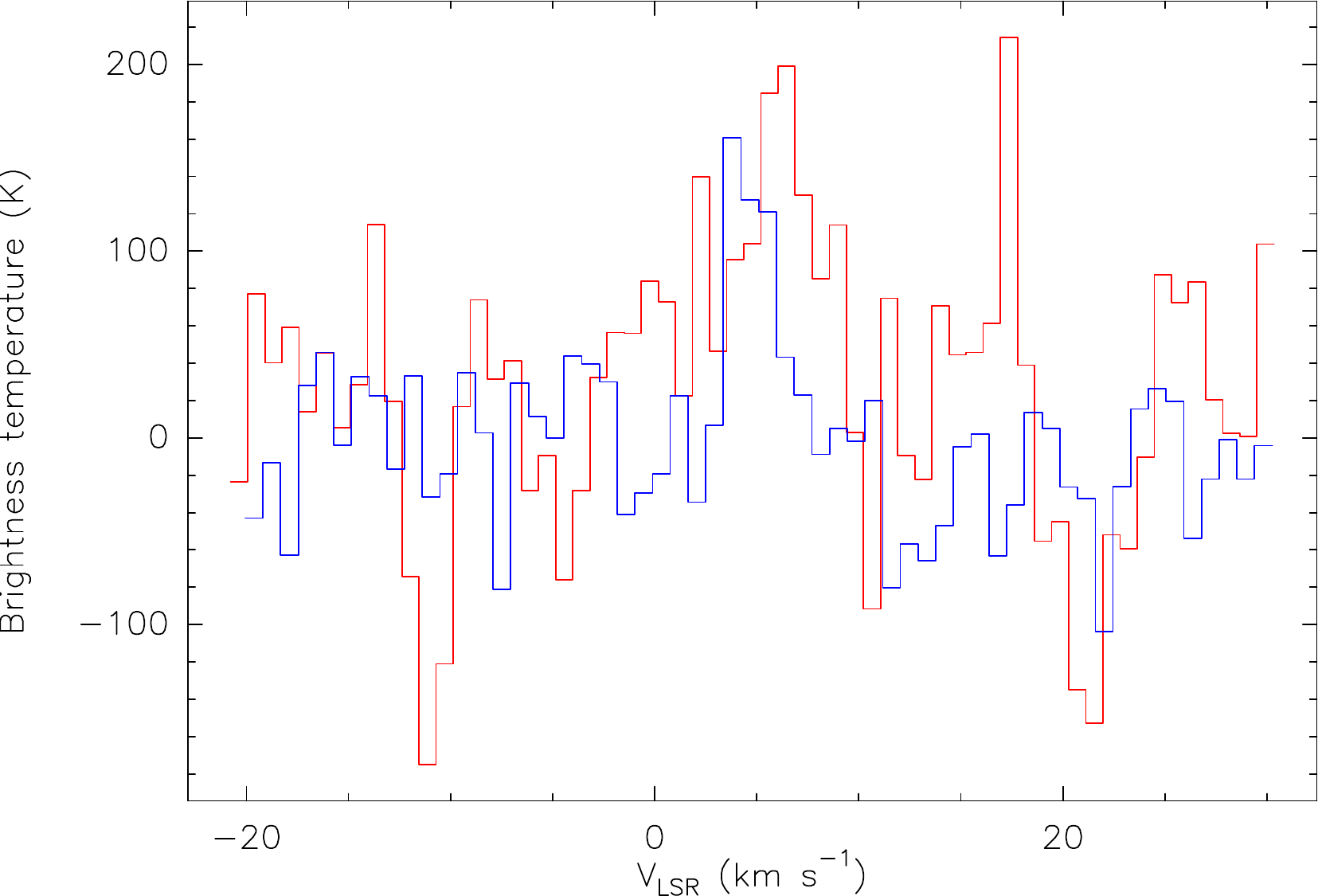}
    \hspace{5mm}
    \includegraphics[width=0.4\textwidth]{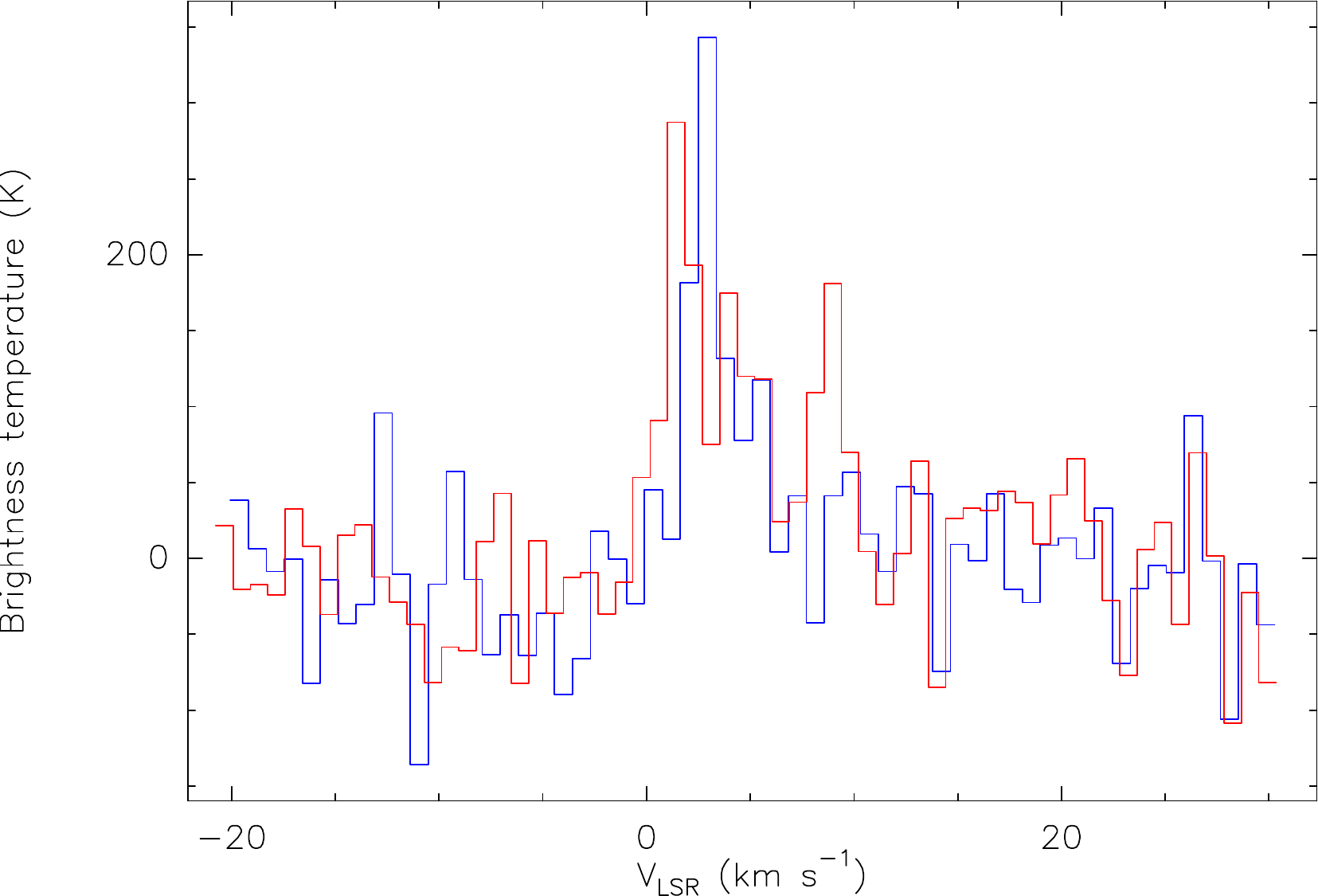}
    \caption{Spectra of the C$^{34}$S(7--6) (blue) and CH$_{3}$CN(19$_3$--18$_3$) (red) emission in the NE (left panel) and SW (right panel) jet lobes in the regions indicated in Figure~\ref{fig:knots_maps}.}
    \label{fig:knots_spectra}
\end{figure*}

In Figure~\ref{fig:knots_spectra} we show the C$^{34}$S(7--6) and CH$_{3}$CN(19$_3$--18$_3$) line spectra in the regions marked in Figure~\ref{fig:knots_maps}. It is worth noting that velocities of the gas emitting in the C$^{34}$S(7--6) and CH$_{3}$CN(19$_3$--18$_3$) lines are somewhat different. In the NE (red-shifted) lobe the CH$_{3}$CN(19$_3$--18$_3$) line is red-shifted relative the C$^{34}$S(7--6) line, while in the SW (blue-shifted) lobe it is blue-shifted. The C$^{34}$S(7--6) velocities in the NE and SW lobes are close to each other. These features hint that the C$^{34}$S(7--6) emission arises mainly in the ambient gas, while the CH$_{3}$CN(19$_3$--18$_3$) emission is also formed in the gas entrained by the jet. The brightness temperature in the lines reaches $\sim$200~K in the NE lobe and $\sim$300~K in the SW lobe. These values represent lower limits for the kinetic temperature of the emitting gas.

\citet{Cesaroni2023, Cesaroni2024} presented and discussed detailed multi-frequency observations of the jet in vicinity of the NIRS3 source. As mentioned above, they detected the NE and SW knots as single entities in contrast to our data (due to a significantly lower angular resolution in their observations). They found that the free-free emission of the ionized gas dominates at frequencies {\la}30~GHz (the highest frequency was about 92~GHz). The fluxes at 92.2~GHz measured in September 2021 are 18.8~mJy for the NE knot and 23.4~mJy for the SW knot. They estimate that the flux of the SW free-free emission is 14.6~mJy after subtraction of the dust contribution, which can be somewhat questionable. Our summary flux for the NE1+NE2 is \iz{$14.8\pm 1.9$~mJy (obtained by summation of the values in Table~\ref{tab:cont})}. This value, within the uncertainties, is consistent with the optically thin free-free emission. However, it is higher than in the model for the free-free emission suggested in \citet{Cesaroni2024}. Our low resolution continuum map (Figure~\ref{fig:cont}) shows a peak at the NE2 position. The flux density in this map integrated in the circle of $\sim$250~mas in diameter around this peak is about 35~mJy. Apparently, it includes contribution of the surrounding dust. The situation with the SW knots is less clear because \citet{Cesaroni2023, Cesaroni2024} report summary fluxes of our central clump and the SW knots. In our data the emission of the SW knots looks very similar to that of the NE knots and we believe that it can be interpreted in the same way.
The emission measure of the knots in our data can be estimated as $EM\sim 6\times 10^{9}$~pc\,cm$^{-6}$, which implies the turnover frequency of $\sim$30~GHz. This is more or less consistent with the spectra presented in \citet{Cesaroni2024}. The electron density is about the same as in the central core, $n_\mathrm{e}\sim \iz{5\times 10^6}$\pcmm.

\subsection{Kinematics of the material in the disk}
The PV diagrams in several lines (Fig.~\ref{fig:pv}) indicate a sub-Keplerian rotation, as found also in \citet{Liu2020}. In fact, the earlier data presented in \citet{Zin15} lead to the same conclusion if the edge-on orientation of the disk is assumed. We do not question the estimates of the mass of the protostar, which was derived from the bolometric luminosity.

The absorption spectra presented in Figure~\ref{fig:absorption} show the deepest feature at $V_\mathrm{LSR}\sim 10$\kms, which is red-shifted in respect to the systemic velocity of this core ($V_\mathrm{LSR}\sim 5$\kms). Therefore, it can be produced by the infalling material. As mentioned in Sect.~\ref{sec:abs}, there is an emission feature at about --5\kms\ in C$^{34}$S(7--6) and CH$_{3}$CN(19$_3$--18$_3$). It could be related to the infalling material behind the core. 

%\subsection{Clumpiness}
%Some models of massive star formation, involved for explaining the accretion bursts, which have been observed in several objects already, exploit fragmentary structure of protostellar cores and disks \citep[e.g.,][]{Meyer17, Meyer19}. There are many implicit indications of small-scale clumpiness in such regions.
%As mentioned above, the image of the diffuse continuum emission in this region looks very clumpy (Fig.~\ref{fig:cont}) but this clumpiness can be an artifact, at least partly. On the other hand, the molecular line emission is very inhomogeneous, indeed, both in space and in velocity (Fig.~\ref{fig:c34s-chmap}). At the same time we do not see large scale coherent structures like in the models mentioned above.

\section{Conclusions}

In this paper we presented observations of the actively investigated high-mass star-forming region S255IR, which harbors the $\sim$20~M\sun\ protostar NIRS3, with the angular resolution of $\sim$15~mas, which corresponds to $\sim$25~au and is almost an order of magnitude better than in the previous studies of this object. The observations were performed with ALMA at 0.9~mm in continuum and in several molecular lines. The main results are as follows.

1. In continuum we detected the central bright source (brightness temperature $\sim$850~K) elongated along the jet direction and two pairs of bright knots in the jet lobes. The distances between the knots in the pairs correspond to the time interval $\sim$1.5~years, taking into account the velocities of the knots determined earlier. This time interval coincides with the interval between the 6.7~GHz maser emission peaks at a certain velocity. It probably implies a double ejection from NIRS3 several years ago.

2. The orientation of the jet differs by $\sim$20$^\circ$ from that on larger scales, as mentioned also in some other recent works. This implies a strong jet precession.

3. The 0.9~mm continuum emission of the central source represents a mixture of the dust thermal emission and free-free emission of the ionized gas. For the ionized gas \iz{assuming an optically thin emission at frequencies \ga22~GHz} we obtain the emission measure of $EM\sim \iz{(0.5-2)}\times 10^{10}$~pc\,cm$^{-6}$ and the electron density of $n_\mathrm{e}\sim\iz{(0.5-1)}\times 10^7$\pcmm. Such properties are typical for hypercompact \Hii\ regions. For a uniform mixture of gas and dust these estimates imply a very low ionization fraction. It is not excluded that dust, at least partly, is expelled from the ionized region and forms a dusty cocoon around it. 
\iz{An alternative model for this source can be based on a hypercompact \Hii\ region with the turnover frequency \ga200~GHz, which explains better the spectral index in the range 92--340~GHz but has a problem with matching the observed brightness temperature. Observations at several frequencies with a sufficiently high resolution are required to select the most appropriate model.}
In the continuum emission of the knots in the jet the free-free component apparently dominates.

4. In the C$^{34}$S(7--6), SiO(8--7) and CO(3--2) lines a rotating disk around NIRS3 about 400~au in diameter is observed. The rotation is sub-Keplerian. There are absorption features in the molecular lines towards the central bright source. The deepest absorption features are red-shifted relative the core velocity, which may indicate an infall. 

5. The molecular line emission \iz{appears} very inhomogeneous at small scales, which \iz{may} indicate a small-scale clumpiness in the disk. 

6. The molecular emission is observed in vicinity of the second (closer to the central source) knots in the pairs of knots in the jet. In the NE (red-shifted) lobe the CH$_{3}$CN(19$_3$--18$_3$) line is red-shifted relative the C$^{34}$S(7--6) line, while in the SW (blue-shifted) lobe it is blue-shifted. The C$^{34}$S(7--6) velocities in the NE and SW lobes are close to each other. These features hint that the C$^{34}$S(7--6) emission arises mainly in the ambient gas, while the CH$_{3}$CN(19$_3$--18$_3$) emission is also formed in the gas entrained by the jet.

\begin{acknowledgements}
This work was supported by the Russian Science Foundation grant number 24-12-00153 (https://rscf.ru/en/project/24-12-00153/).
We are grateful to Stan Kurtz for the helpful discussions \iz{and to the anonymous referee for the useful comments}.
This paper makes use of the following ALMA data: ADS/JAO.ALMA \#2019.1.00315.S. ALMA is a partnership of ESO (representing its member states), NSF (USA), and NINS (Japan), together with NRC (Canada), MoST and ASIAA (Taiwan), and KASI (Republic of Korea), in cooperation with the Republic of Chile. The Joint ALMA Observatory is operated by ESO, AUI/NRAO, and NAOJ
\end{acknowledgements}



\bibliographystyle{aa}
\bibliography{s255ir} % if your bibtex file is called example.bib

%\begin{thebibliography}{99}
%\end{thebibliography}

%%%%%%%%%%%%%%%%%%%%%%%%%%%%%%%%%%%%%%%%%%%%%%%%%%

%%%%%%%%%%%%%%%%% APPENDICES %%%%%%%%%%%%%%%%%%%%%

%\appendix


%If you want to present additional material which would interrupt the flow of the main paper, it can be placed in an Appendix which appears after the list of references.

%%%%%%%%%%%%%%%%%%%%%%%%%%%%%%%%%%%%%%%%%%%%%%%%%%

% Don't change these lines
%\bsp	% typesetting comment
%\label{lastpage}
\end{document}